\documentclass[twocolumn,trackchanges]{aastex701}
\usepackage[caption=false]{subfig}
\usepackage{amsmath}

\shortauthors{Christy et al.}
\shorttitle{Radio Polarimetry of GRB\,260310A}
\begin{document}
\title{First Detection of Faraday Rotation in a Gamma-Ray Burst Afterglow: Low Polarization and High Rotation Measure in GRB 260310A Reveal Jet Magnetic Structure and Environment}

\newcommand{\UA}{\affiliation{Steward Observatory, University of Arizona, 933 North Cherry Avenue, Tucson, AZ 85721-0065, USA}}
\author[orcid=0000-0003-0528-202X,gname=Collin,sname=Christy]{Collin~T.~Christy}
\UA
\email[show]{collinchristy@arizona.edu}

\author[orcid=0000-0003-1792-2338,gname=Tanmoy,sname=Laskar]{Tanmoy~Laskar}
\affiliation{Department of Physics \& Astronomy, University of Utah, Salt Lake City, UT 84112, USA}
\email{tanmoy.laskar@utah.edu}

\author[orcid=0000-0002-8297-2473,gname=Kate,sname=Alexander]{Kate~D.~Alexander}
\UA
\email{kdalexander@arizona.edu}

\author[orcid=0000-0003-4537-3575,gname=Noah,sname=Franz]{Noah~Franz}
\email{nfranz@arizona.edu}
\UA

\author[orcid=0000-0001-8530-8941, gname=Jonathan, sname=Granot]{Jonathan Granot}
\affiliation{Department of Natural Sciences, The Open University of Israel, P.O Box 808, Ra'anana 4353701, Israel}
\affiliation{Astrophysics Research Center of the Open University (ARCO), The Open University of Israel, P.O Box 808, Ra'anana 43537, Israel}
\affiliation{Department of Physics, The George Washington University, 725 21st Street NW., Washington, DC 20052, USA}
\email{granot@openu.ac.il}

\author[orcid=0000-0002-7706-5668, gname=Ryan, sname=Chornock]{Ryan Chornock}
\affiliation{Department of Astronomy, University of California, Berkeley, CA 94720-3411, USA}
\affiliation{Berkeley Center for Multi-messenger Research on Astrophysical Transients and Outreach (Multi-RAPTOR), University of California, Berkeley, CA 94720-3411, USA}
\email{chornock@berkeley.edu}

\author[orcid=0000-0003-4768-7586, gname=Raffaella, sname=Margutti]{Raffaella Margutti}
\affiliation{Department of Astronomy, University of California, Berkeley, CA 94720-3411, USA}
\affiliation{Berkeley Center for Multi-messenger Research on Astrophysical Transients and Outreach (Multi-RAPTOR), University of California, Berkeley, CA 94720-3411, USA}
\affiliation{Department of Physics, University of California, 366 Physics North MC 7300, Berkeley, CA 94720, USA}
\email{rmargutti@berkeley.edu}

\author[orcid=0000-0003-0516-2968, gname=Ramandeep, sname=Gill]{Ramandeep Gill}
\affiliation{Astrophysics Research Center of the Open University (ARCO), The Open University of Israel, P.O Box 808, Ra'anana 43537, Israel}
\affiliation{Instituto de Radioastronom\'ia y Astrof\'isica, Universidad Nacional Aut\'onoma de M\'exico, Morelia, Michoac\'an, C.P. 58089, M\'exico}
\email{rsgill.rg@gmail.com}

\author[orcid=0000-0002-0744-0047, gname=Jeniveve, sname=Pearson]{Jeniveve Pearson}
\UA
\email{jenivevepearson@arizona.edu}

\author[orcid=0000-0002-9392-9681, gname=Edo, sname=Berger]{Edo Berger}
\affiliation{Center for Astrophysics $\vert$ Harvard \& Smithsonian, Cambridge, MA 02138, USA}
\email{eberger@cfa.harvard.edu}

\author[0000-0002-7374-935X, gname=Wen-fai, sname=Fong]{Wen-fai Fong}
\email{wfong@northwestern.edu}
\affiliation{Center for Interdisciplinary Exploration and Research in Astrophysics and Department of Physics and Astronomy, \\Northwestern University, 1800 Sherman Ave., 8th Floor, Evanston, IL 60201, USA}
\affiliation{Department of Physics and Astronomy, Northwestern University, Evanston, IL 60208, USA}

\author[orcid=0009-0007-3919-8439, sname='Rohde']{Coleman Rohde}
\affiliation{Department of Physics \& Astronomy, University of Utah, Salt Lake City, UT 84112, USA}
\email{rohdecoleman@google.com}  

\author[orcid=0000-0002-1214-770X, gname=Patricia, sname=Schady]{Patricia Schady}
\affiliation{Department of Physics, University of Bath, Claverton Down, Bath
BA2 7AY, UK}
\email{ps2018@bath.ac.uk}

\begin{abstract}
We report the detection of linear polarization in the radio afterglow of GRB 260310A, representing the first centimeter-wavelength polarization detection of a gamma-ray burst (GRB) afterglow and the first measurement of Faraday rotation in a GRB environment. We detect linearly polarized emission across $11-25$ GHz, with a polarization fraction decreasing monotonically from $(3.18 \pm 0.18)\%$ at 25 GHz to $(0.69 \pm 0.22)\%$ at 11 GHz. Interpreting the radio data as emission from a reverse shock in a structured, relativistic jet, the observed depolarization toward lower frequencies is consistent with suppression by synchrotron self-absorption, while the low observed polarization at high frequencies relative to the theoretical maximum suggests a patchy magnetic field in the jet with a coherence scale, $\theta_{\rm B}\approx10^{-2}$\,rad. We identify a frequency-dependent rotation of the polarization angle consistent with Faraday rotation, with a rotation measure of ${\rm RM} = -(8300 \pm 90)~\rm{rad/m^2}$ at the GRB redshift. The magnitude of the rotation measure is consistent with propagation through a dense, magnetized environment, such as a progenitor HII region. These findings demonstrate that GRB afterglows exhibit measurable linear polarization at centimeter wavelengths, and that their polarimetric properties probe both intrinsic jet magnetization and the surrounding medium. Future multi-frequency polarimetric monitoring over timescales of days to weeks will enable detailed studies of the evolution of magnetic field structure and provide new constraints on the role of magnetic fields in GRB afterglows.
\end{abstract}

\section{Introduction}
Magnetized relativistic jets are ubiquitous in extreme accretion environments across a range of mass scales, from stellar-mass compact objects to supermassive black holes (SMBHs) in active galactic nuclei (AGN). However, in spite of their prevalence, the fundamental physics governing their formation, collimation, and magnetization remain poorly understood \citep{m06, kz15}. Long-duration gamma-ray bursts (GRBs) resulting from the core collapse of massive stars offer a unique laboratory for addressing these key questions, as their extreme luminosity and large bulk Lorentz factors ($\Gamma\sim10^2-10^3$) make them ideal probes for jet physics observable on human timescales \citep{mr93,rm94,fkn97,wb06}.

Synchrotron polarization from GRB afterglows offers a window into the magnetic field geometry of jetted systems. The afterglow itself is generated by the interaction of the jet with the immediate circumburst environment, producing two shocks: a long-lived relativistic forward shock (FS) in the ambient medium powering the broadband afterglow, and a short-lived reverse shock (RS) propagating back through the decelerating ejecta \citep{sp95, kz03a}. Both of these shocks accelerate electrons and generate synchrotron emission, whose polarization directly encodes the magnetic field geometry. Ordered fields are expected to yield high RS linear polarization ($\Pi_{\rm RS} \gtrsim 70\%$), while tangled fields produce low polarization ($\Pi_{\rm RS} \lesssim 10\%$) that depends on the anisotropy of the magnetic fields \citep{gk03, gt05, gra03, gw99}. FS linear polarization, meanwhile, is sensitive to jet angular structure and viewing geometry through its behavior around the jet break time, providing complementary constraints on whether the jet has, e.g., a top-hat, power-law, or Gaussian energy profile \citep{sal03, rlsg04, bgb+26}. 

Multi-frequency polarization observations can provide additional information via the detection of Faraday rotation, which is observed regularly in AGN \citep[e.g., ][]{hla12, gkr15, kks17}, Fast Radio Bursts \citep[e.g.,][]{2018Natur.553..182M, 2023MNRAS.520.2039Y}, and Galactic sources \citep[e.g.,][]{rk89,os93,hfm04}. Faraday rotation, which manifests as a rotation of the polarization position angle with frequency, directly constrains the line-of-sight (LOS) magnetic field strength and electron density of the intervening medium \citep{b66}. This provides a direct probe of the magnetic field conditions intrinsic to the afterglow emitting region or the intervening circumburst and host galaxy environments and may constrain GRB progenitor models \citep{gt05,gkg21}. 

Optical and near-infrared polarimetric observations of GRB afterglows have yielded successful detections in a handful of events \citep[e.g.,][]{clg99, rwv00, bhp02, bmg03, bsc03, lcd03, mmp06, mkca+13, wct+14,cg16, ska17,arimoto+24}. In most cases, the earliest ($<10^3$ s) and most highly polarized ($10\% \lesssim  \Pi_L \lesssim 30\%$) detections are likely dominated by RS emission \citep{gkg21}. As the optical emission becomes dominated by FS emission, the intrinsic polarized RS emission can be diluted by the weakly polarized FS component, as proposed for GRB 120308A \citep{mkca+13}.

As RS emission persists longest at low frequencies, it is natural to search for RS polarization signatures in the radio and millimeter. Historically, radio polarization searches in GRBs have mostly yielded upper limits \citep[e.g.,][]{gt05, vdhpdb+14, utc+23}, in part due to the limited sensitivity of millimeter/centimeter-band facilities and the large cosmological distances of most events. This changed with the first ALMA detection of linearly polarized emission ($\approx 1\%$) from the nearby ($z=0.4245$) GRB~190114C \citep{lag+19}, which demonstrated that GRB reverse shocks are polarized at mm wavelengths. These observations constrained the magnetic field coherence scale in the ejecta, inferred in this case to be $\theta_B\approx10^{-3}\,$rad, which is at least $\sim$10 times smaller than suggested by the large $28\pm4\,$\% polarization of the early 240\,--\,323\;s optical afterglow of GRB\,120308A \citep{mkca+13}, showing significant diversity in the GRB population. However, GRB 190114C's afterglow faded quickly, enabling only one constraining single-band polarimetry observation within hours of the GRB discovery. 

Here we present radio polarimetry observations of the recent GRB 260310A, which, due to its proximity, has one of the brightest radio afterglows seen in decades. We report the first multi-frequency detection of linear polarization in a GRB afterglow, extending radio polarimetry into the centimeter-band and opening a new observational window into GRB jet magnetization and structure. Additionally, the position angle exhibits Faraday rotation with frequency, marking the first detection of this effect in a GRB and revealing information about the GRB's environment. 

We structure our paper as follows. In section \ref{sec:obs}, we summarize basic properties of GRB 260310A, compile relevant public multi-wavelength observations from the community, and present our radio observations and data reduction procedure. We model the multi-wavelength photometry and the radio polarization in Section \ref{sec:mod} and interpret these findings jointly in Section \ref{sec:disc}. We conclude in Section \ref{sec:conc}. Unless otherwise specified, frequencies and times are in the observer frame. For our analysis, we assume a cosmology of $\Omega_M=0.31$, $\Omega_{\Lambda}=0.69$, $h=0.68$, and a luminosity distance at $z=0.153$ of 745 Mpc. %$2.3\times10^{27}$\,cm.

\section{Observations}\label{sec:obs}

\begin{figure}
    \centering
    \includegraphics[width=\linewidth]{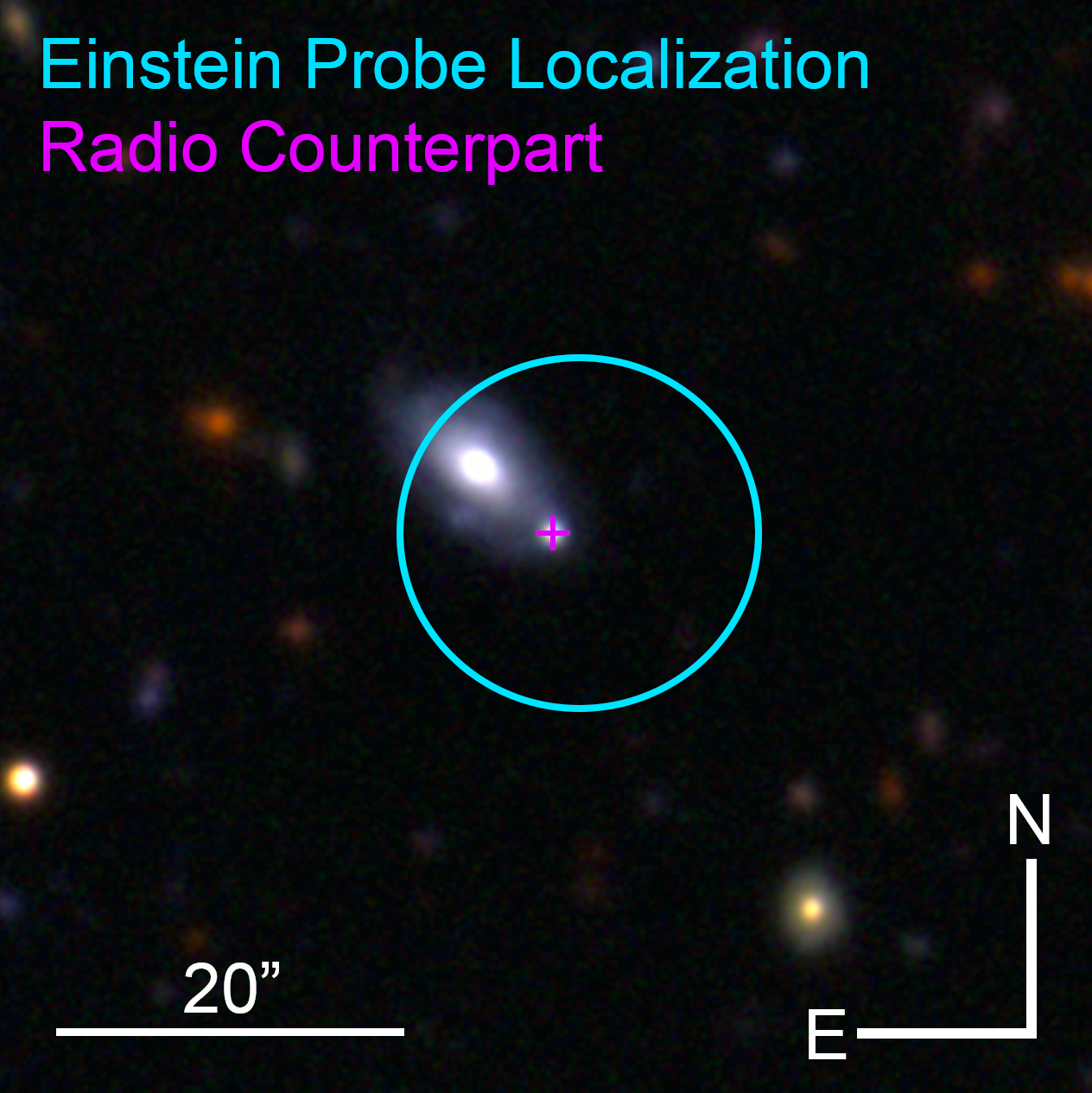}
    \caption{MMT Binospec $gri$ color image of GRB~260310A taken on 2026 Apr 18, $\sim 39.2$ d post trigger. The coordinates of our VLA detection are shown as a purple cross and are consistent with the optical position. The statistical error on the radio position is smaller than the systematic astrometric accuracy of $\sim10~\rm{mas}$ typically achievable by the VLA in A configuration. For comparison, the Einstein Probe 90\% localization region of the GRB is shown as a blue circle \citep{gcn43994}.}
    \label{fig:host}
\end{figure}

\subsection{Basic properties and community data: GCNs}
GRB 260310A triggered the Fermi Gamma-ray Burst Monitor (GBM) at 04:57:10 UT on 10 March 2026 \citep{gcn43951}; all times reported in our subsequent analysis are measured relative to this time. The prompt gamma-ray emission consists of a single primary pulse with $T_{90}\sim60$ sec \citep[$50-300$ keV;][]{gcn43975} and was also detected by AstroSAT CZTI, consistent with a standard long GRB \citep{gcn43958,gcn44053}. The afterglow was subsequently detected in the optical \citep{tns_discovery,hhp+26,gcn43974}, X-rays \citep{gcn43994}, and radio \citep{gcn44005,gcn44057,gcn44045,gcn44235,gcn44160}. Shortly after, a spectroscopic redshift of $z=0.153$ was reported and confirmed by multiple groups \citep{gcn43984,gcn43977,gcn43986}. The transient is located in the outskirts of its host galaxy with a separation from the nucleus of $\approx 5.9''$ ($\sim 20$ kpc; see Figure \ref{fig:host}). At this redshift, the peak energy, $E_{\rm peak}=(172\pm5)$\,keV, cut-off power-law index $=5.8\pm0.4$, and fluence $(5.8\pm0.4)\times10^{-6}\,{\rm erg\,cm^{-2}}$ (10--1000\,keV, observer frame; \citealt{gcn43975}) correspond to an isotropic equivalent $\gamma$-ray energy of $E_{\gamma,\rm iso}=(3.4\pm0.2)\times10^{50}$\,erg (1--$10^4$\,keV, rest frame). These energetics place GRB 260310A in the compact object merger region of the $E_{p,i}-E_{\text{iso}}$ relation \citep{a06,gcn43981}, which is unusual but not unprecedented for long-duration GRBs (e.g., GRB 211211A; \citealt{rgl22,tfo22}). Continued spectroscopic follow-up of this event revealed an emerging Ic-BL supernova component, SN 2026fgk \citep{gcn44124, gcn44125, gcn44137, gcn44248}, securing a connection to the collapse of a massive star.

To provide a broader multi-wavelength context for our polarization observations (Section \ref{sec:vla}), we utilize a subset of the public radio, optical, and X-ray photometry\footnote{The X-ray light curve exhibits a re-brightening at $\gtrsim$ \citep{gcn44234,gcn44375,gcn44278}, which cannot be explained with a standard afterglow model. We defer a discussion of these data to future work.} reported by the community in GCN circulars  \citep{gcn43974,gcn43979,gcn43980,gcn43990,gcn43991,gcn43993,gcn43994,gcn43996,gcn44000,gcn44001,gcn44002,gcn44003,gcn44004,gcn44005,gcn44006,gcn44021,gcn44045,gcn44057,gcn44058,gcn44060,gcn44063,gcn44067,gcn44073,gcn44095,gcn44096,gcn44104,gcn44112,gcn44134,gcn44160,gcn44235}. We also use optical polarimetric observations conducted by \citet{gcn43990} at 2.89 d post-trigger, which resulted in a linear polarization upper limit of $\Pi_L<1.5$\% ($3\sigma$, not corrected for Galactic interstellar polarization). 
We exclude optical photometry at $t>15.2$ days as these observations deviate from the power-law decline, indicating the emergence of an underlying supernova \cite{gcn44125}.

\subsection{VLA polarization observations}\label{sec:vla}

\begin{figure*}
    \centering    \includegraphics[width = \textwidth]{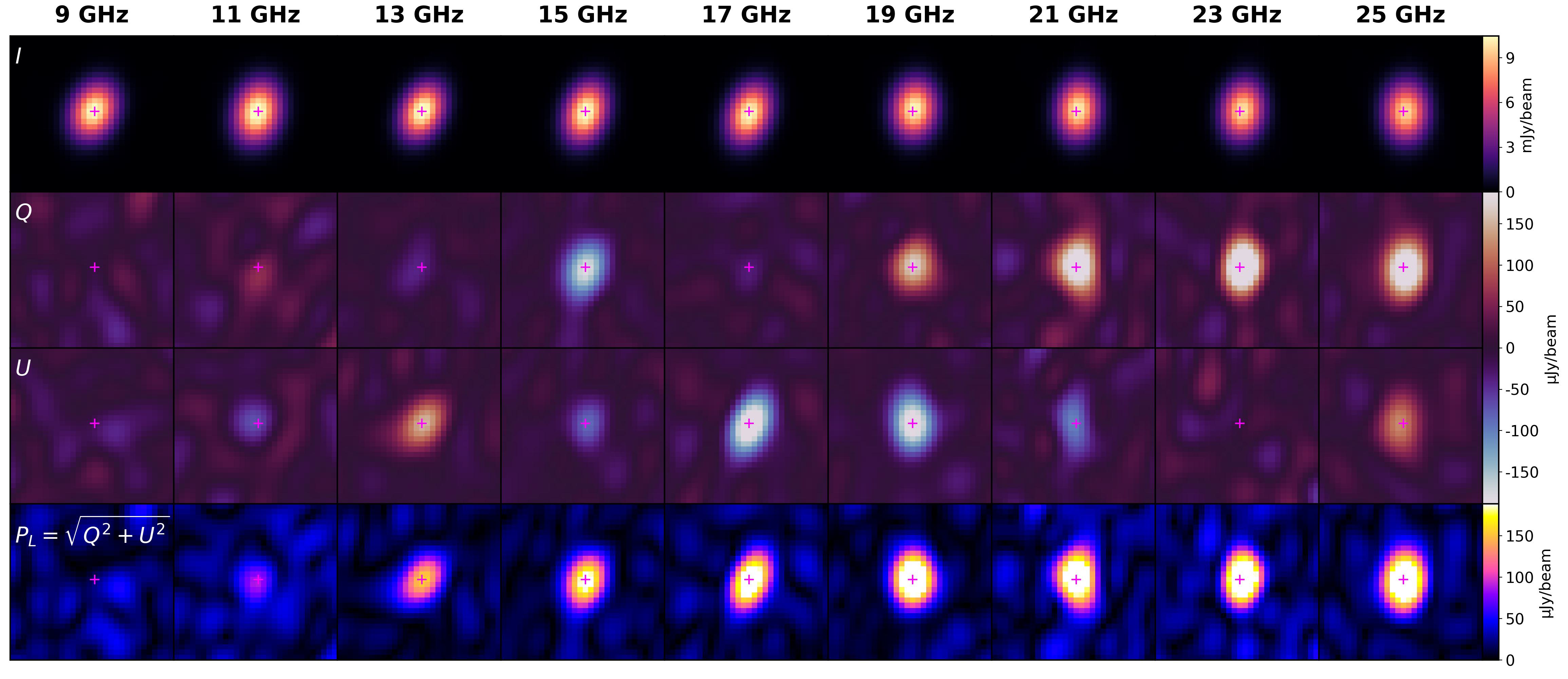}
    \caption{Stokes $IQU$ and linear polarization intensity $P_L=\sqrt{Q^2+U^2}$ of GRB 260310A at $19.2~\rm{d}$. The images reveal a strong polarized point source that fades at lower frequencies. Additionally the $QU$ images reveal a frequency-dependent rotation in the polarization angle.}
    \label{fig:pol_imgs}
\end{figure*}

On 2026 Mar 29 ($19.2~\rm{d}$), we observed GRB 260310A ($\alpha=14^\text{h}37^\text{m}16^\text{s}.14$, $\delta=+71^\circ50'30''.30$; J2000) with the Karl G. Jansky Very Large Array (VLA) under program VLA/26A-032 (PI: T. Laskar). We performed a 3 hr continuum observation using 3-bit samplers at bands X (10 GHz), Ku (15 GHz), and K (22 GHz). We observed 3C286 as the bandpass, flux, and polarization angle calibrator, J1407+2827 as a low-polarization leakage calibrator, and J1357+7643 as the complex gain calibrator.

We processed the data with the Common Astronomy Software Application (CASA 6.7.0.31; \citealt{CASA_team}). We first calibrated the parallel-hand (RR/LL) cross-correlation visibilities using the NRAO pipeline (v.\ 6.4.1.12). We then re-applied the pipeline calibration solutions with parallactic angle corrections disabled. For the polarization calibration, we first corrected for the instrumental delay between the cross-hands (RL/LR) using the bandpass calibrator separately for each baseband. We then solved for the instrumental frequency-dependent leakage terms (D-terms) using the low-polarization source J1407+2827. Finally, we solved for the frequency-dependent R-L phase per antenna using a polarization model of 3C286 constructed from polynomial fits to the tabulated flux densities, polarization fractions, and polarization angles provided by NRAO\footnote{\url{https://science.nrao.edu/facilities/vla/docs/manuals/obsguide/modes/flux-density-scale-polarization-leakage-polarization-angle-tables}}.

For the target field, we performed phase-only self-calibration on Stokes $I$, with the derived phase solutions subsequently applied to all Stokes parameters. The effect on the measured flux density was modest ($\lesssim5\%$ across all bands), with the largest corrections occurring at the highest frequencies. We imaged the data using the CASA task \verb'tclean', and obtained flux densities and their uncertainties by using the CASA task \verb'imfit' to fit an elliptical Gaussian fixed to the size of the synthesized beam. For estimating Stokes $Q$ and $U$ flux densities, we fixed the position of the elliptical Gaussian to the source location measured in the Stokes $I$ image. We discuss additional calibration checks performed on the data in Appendix \ref{sec:polsys}. We present the $I$, $Q$, $U$, and $P_L=\sqrt{Q^2+U^2}$ Stokes images in Figure \ref{fig:pol_imgs}.

\begin{deluxetable*}{c|ccccccc} % Specify column alignment
\tablecaption{Measured Stokes flux densities. Uncertainties are $1\sigma$ and statistical only, upper limits are $3\sigma$.\label{tab:Stokes_pol}}
\tablehead{\colhead{$\nu_{\rm{obs}}$ (GHz)} & \colhead{I (mJy)} &\colhead{Q ($\rm{\mu}Jy$)} & \colhead{U ($\rm{\mu}Jy$)}& \colhead{ $P_L$ ($\rm{\mu}Jy$)} & \colhead{$\Pi_L$ ($\%$)} & \colhead{$\chi$ (rad)}}
\startdata
9 & 10.21 $\pm$ 0.02 & 0 $\pm$ 16 & -27 $\pm$ 13 & 27 $\pm$ 21 & $<0.6$ &- \\
11 & 10.54 $\pm$ 0.02 & 56 $\pm$ 17 & -47 $\pm$ 16 & 73 $\pm$ 23 & 0.69 $\pm$ 0.22 & -3.49 $\pm$ 0.11 \\
13 & 10.49 $\pm$ 0.01 & -42 $\pm$ 9 & 163 $\pm$ 11 & 169 $\pm$ 14 & 1.61 $\pm$ 0.13 & -2.23 $\pm$ 0.03 \\
15 & 10.39 $\pm$ 0.01 & -191 $\pm$ 9 & -80 $\pm$ 10 & 207 $\pm$ 13 & 1.99 $\pm$ 0.13 & -1.37 $\pm$ 0.02  \\
17 & 10.23 $\pm$ 0.01 & -27 $\pm$ 11 & -241 $\pm$ 13 & 243 $\pm$ 17 & 2.37 $\pm$ 0.17 & -0.84 $\pm$ 0.02 \\
19 & 9.98 $\pm$ 0.01 & 170 $\pm$ 10 & -221 $\pm$ 9 & 279 $\pm$ 14 & 2.79 $\pm$ 0.14 & -0.46 $\pm$ 0.02 \\
21 & 9.77 $\pm$ 0.03 & 251 $\pm$ 17 & -103 $\pm$ 14 & 271 $\pm$ 22 & 2.77 $\pm$ 0.22 & -0.20 $\pm$ 0.03\\
23 & 9.54 $\pm$ 0.03 & 285 $\pm$ 17 & -3 $\pm$ 17 & 285 $\pm$ 24 & 2.99 $\pm$ 0.25 & -0.01 $\pm$ 0.03 \\
25 & 9.24 $\pm$ 0.02 & 261 $\pm$ 12 & 135 $\pm$ 12 & 294 $\pm$ 17 & 3.18 $\pm$ 0.18 & 0.24 $\pm$ 0.02 \\
\enddata
% \tablecomments{Notes if needed here.}
\end{deluxetable*}

We detect linearly polarized emission from GRB 260310A from $11-25$ GHz (Table \ref{tab:Stokes_pol}). The polarization fraction drops slowly from $(3.18\pm0.18)\%$ at $25$\,GHz to $(1.99\pm0.13)\%$ at 15\,GHz, and then plummets to undetectable values ($<0.6\%$, $3\sigma$) at 9\,GHz (Figure \ref{fig:RMQU}, left panels). Our lowest significant polarization detection of $(0.69\pm0.22)\%$ is at 11\,GHz, which coincides with the peak of the Stokes $I$ spectrum.

While our observations at $19.2$~days do not fully sample the low-frequency segment, VLA observations at $17.2$~days reveal a steep low-frequency spectrum\footnote{We use the convention $F_\nu\propto t^\alpha\nu^\beta$ throughout.} with $\beta=1.37\pm0.02$ \citep{gcn44160}. This steeply rising spectrum indicates that the emission likely experiences strong synchrotron self-absorption (SSA). Assuming the spectral peak is the SSA break ($\nu_{\rm sa}$), we fit the observed depolarization as a function of frequency (see Appendix~\ref{appendix:PiSSA}, Equation\,\ref{eq:fracpolofSR}) with the intrinsic linear polarization on the optically thin segment ($\Pi_0$) as a free parameter. We find that this model reproduces the $\Pi_L(\nu)$ spectrum well for $\Pi_0=(3.2\pm0.2)\%$ and electron energy index, $p=2.7^{+0.8}_{-0.5}$, and predicts $\nu_{\rm sa}=(13.3\pm0.5)$\,GHz, roughly consistent with the peak of the Stokes $I$ spectrum, and providing further evidence that the observed depolarization is attributable to SSA. 

We find no evidence of temporal evolution in the measured polarized intensities during our observations. However, given the relatively short time on source ($\sim$10, 25, and 45 minutes in X, Ku, and K bands respectively) at $\sim19$ days post-burst these observations span a very small fraction of the afterglow age ($\Delta t/t\lesssim 0.001$). We therefore cannot meaningfully constrain whether the polarization fraction or position angle evolve with time.

\begin{figure*}
   \includegraphics[width=\textwidth]{combined_pol_figure.pdf}
\caption{\textit{(a)} The measured flux density of Stokes $I$, \textit{(b)} the linear polarization flux density, and \textit{(c)} the linear polarization fraction $\Pi_L$ as a function of frequency. We show the best-fitting model (Equation \ref{eq:fracpolofSR}; dotted lines, shaded regions are $1\sigma$ posteriors), which show a constant linear polarization intensity and fraction that decays exponentially to lower frequencies. \textit{(d)} We show the Stokes $U$ vs. $Q$ colorized by observed frequency. The data exhibit a spiral structure indicative of Faraday Rotation. The solid line shows the depolarization frequency response derived from Equation \ref{eq:fracpolofSR}, and the shaded region indicates the average $3\sigma$ detection threshold for $P_L$. \textit{(e)} The polarization angle $\chi = (1/2)\arctan(U/Q)$ as a function of $\lambda^2$. The data reveal a steep slope corresponding to a rotation measure of $|\rm{RM}|\approx6.3\times10^3~\rm{rad/m^2}$ indicative of a highly ordered magnetic field structure along the line of sight.}
\label{fig:RMQU}
\end{figure*}

\subsection{Faraday Rotation}
In Figure~\ref{fig:RMQU} we show the variation of the Stokes $Q$ and $U$ flux densities as a function of frequency. The polarized intensity and position angle, $\chi=\frac12\arctan(U/Q)$, exhibit a spiral pattern characteristic of Faraday rotation combined with frequency-dependent depolarization. The rate at which the polarization position angle traverses this spiral yields the rotation measure (RM), 
\begin{align}
    \chi = \chi_0 + {\rm RM}\,\lambda^2.
\end{align}
After removing $2\pi$ wraps, we find that a plot of $\chi$ vs $\lambda^2$ (Figure~\ref{fig:RMQU}, lower right panel) is well-fit ($R^2=0.999$) with a straight line, yielding\footnote{For this RM, the fractional depolarization due to bandwidth averaging,  $\frac{1}{\Delta\nu}|\int_{\nu_0}^{\nu_0+\Delta\nu}e^{2i\chi(\nu)}d\nu|$ is small ($\lesssim10\%$) over a 2\,GHz bandwidth at $\gtrsim13$\,GHz. This can rise to $\approx25\%$ at $\approx11$\,GHz; however, we find that splitting the bandwidth into smaller chunks and reimaging does not significantly change the shape of $\chi(\nu)$, indicating that the depolarization observed at low frequencies is intrinsic and not due to bandwidth effects.} $\rm{RM}=-(6250\pm70)~\rm{rad/m^2}$.

To determine whether this observed RM is intrinsic to the source, we estimate the expected Faraday rotation from the Milky Way (MW) and the intergalactic medium (IGM). GRB~260310A is located at Galactic coordinates, $l = 111.99^\circ,~b = 42.97^\circ$ and, at these intermediate galactic latitudes, all-sky maps of Galactic Faraday depths suggest that the MW contribution to the RM $\rm{RM_{MW,GRB}}=-(1.12 \pm 8.15)~\rm{rad/m^2}$ is very modest \citep{ojr12}. 

Our data reduction process used the known $\chi(\nu)$ for 3C286 to calibrate $\chi$ for the target. This process corrects for the $\rm{RM_{MW}}$ contribution along the 3C286 LOS, $\text{RM}_{\rm MW,3C286} = (2.52 \pm 2.77)~\rm{rad/m^2}$ \citep{ojr12}. Since the 3C286 LOS ($l = 56.52^\circ,~b = 80.67^\circ$) is separated from that of GRB 260310A by $\sim 42^\circ$, the two probe independent regions of the Galactic foreground. Even accounting for this difference, however, we find that the net systematic $\rm{RM_{MW}}$ offset for the GRB LOS is negligible, $\text{RM}_{\rm MW,GRB} - \text{RM}_{\rm MW,3C286} \approx -(3.6 \pm8.6)~\rm{rad/m^2}$ ($\lesssim$0.1\% of the observed RM). Similarly, the Faraday rotation contribution from the IGM is expected to be small, with current estimates yielding $|\rm{RM_{IGM}}| \lesssim 10~\rm{rad/m^2}$ \citep[e.g.][]{v75,ksn76, vgr19, mdd24}. The combined Galactic foreground and IGM contribution along the GRB LOS is therefore of order $\sim10~\rm{rad/m^2}$.

We further verified this by examining the RM of the phase calibrator J1357+7643 (Figure~\ref{fig:cal_seds} in Appendix~\ref{sec:polsys}), which exhibits a modest $|\rm{RM}|\approx60~\rm{rad/m^2}$. Given the small angular separation between the phase calibrator and GRB 260310A ($\sim5^\circ$), both sight lines probe a similar region of the Galactic foreground and IGM. Comparing to the total MW+IGM contribution computed above would imply that the bulk of the phases calibrator's RM is intrinsic to the source itself, but even if we assume the full $\rm{RM}\approx-60~\rm{rad/m^2}$ is due to the MW and IGM along the LOS to the phase calibrator, this is still two orders of magnitude below our observed GRB $|\rm{RM}|\approx6.3\times10^3~\rm{rad/m^2}$. We therefore conclude that the observed RM of the GRB polarization originates from a Faraday screen local to GRB 260310A or its host galaxy. This implies an intrinsic $\rm{RM_{ int}} = (1+z)^2~RM \approx -8.3\times10^3~\rm{rad/m^2}$ at the source redshift.

\section{Afterglow Modeling}\label{sec:mod}

For a physical interpretation of our polarimetric measurements, we now discuss the multi-wavelength afterglow emission in the standard formalism, in which the RS and FS produce synchrotron radiation via relativistic electrons accelerated to a power-law distribution in energy (with index, $p$). We assume a fraction $\epsilon_e$ of the post-shock energy is imparted to relativistic electrons and a fraction $\epsilon_B$ to magnetic fields. The isotropic-equivalent kinetic energy of the jet (with collimation angle, $\theta_{\rm jet}$) is $E_{\rm K,iso}$. We assume the circumstellar medium (CSM) has a power-law density profile, $\rho=AR^{-k}$, which, can be cast in terms of the parameter $A_*$ as $A=5\times10^{11}A_*\,{\rm g\,cm^{-1}}$ for a wind-like CSM ($k=2$). The spectrum of each emission component (RS and FS) is characterized by \citep[e.g.,][]{sar97,sp99,gs02} a self-absorption break ($\nu_{\rm sa}$), a characteristic frequency ($\nu_m$), and a synchrotron cooling break ($\nu_c$).

\begin{figure*}
  \includegraphics[width=0.33\linewidth]{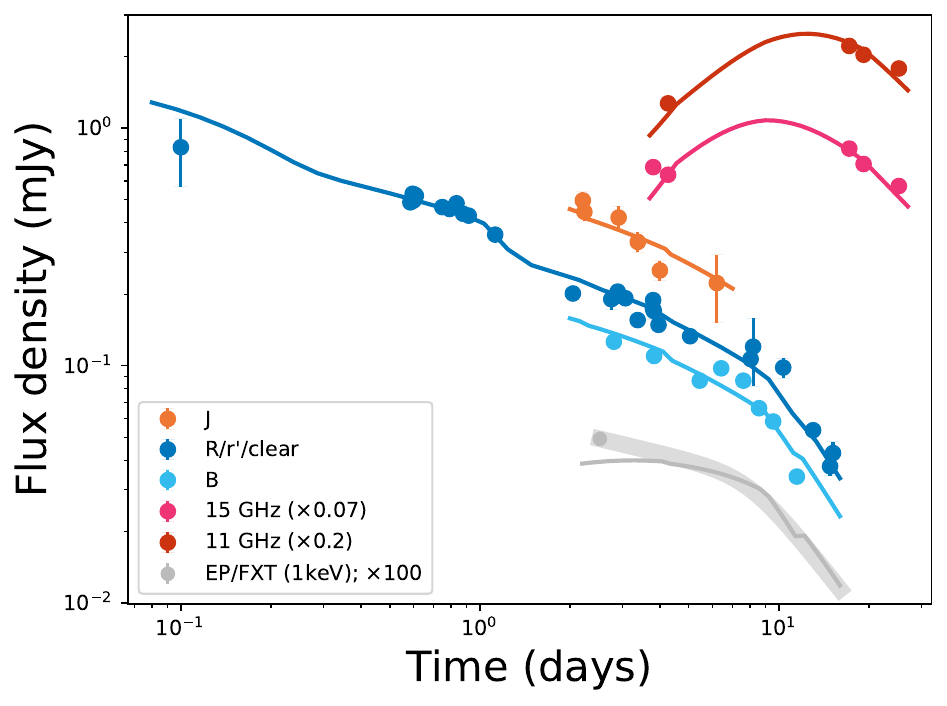}
    \includegraphics[width=0.33\linewidth]{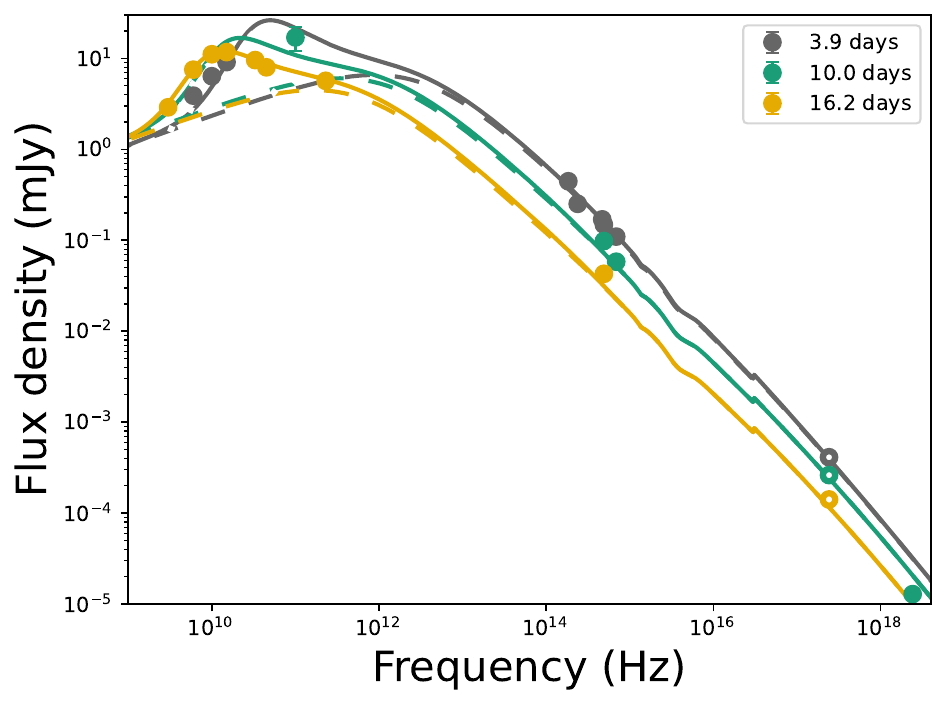}
  \includegraphics[width=0.33\linewidth]{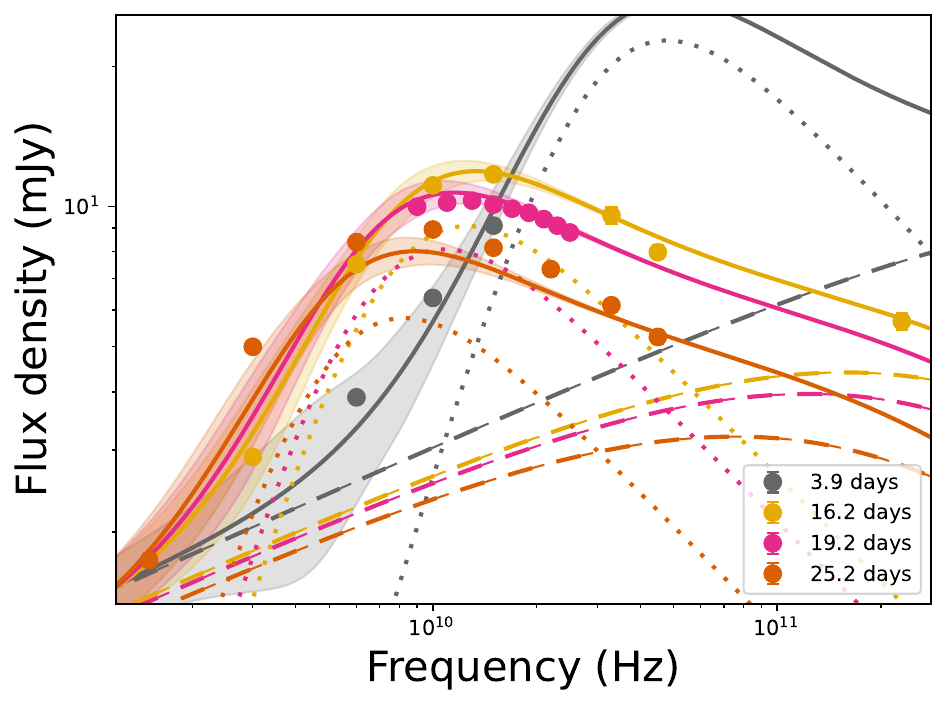}
\caption{Multi-wavelength radio to X-ray light curves at $\approx0.1$--25\,days (left), and spectral energy distributions (center; extrapolated EP/FXT points shown as open symbols) with a zoom in to the radio (right) for GRB\,260310A with a FS model in a wind environment including two episodes of energy injection (dominating in the optical and X-rays; dashed) and a RS in a structured jet (dominating at radio wavelengths; dotted, right panel). The shaded regions indicate the $1\sigma$ scatter expected from interstellar scintillation along this sight line.}
\label{fig:model-FSRS}
\end{figure*}

\subsection{Afterglow Evolution -- Optical \& X-rays}
We find that the optical and X-ray light curves are challenging to interpret in this framework. The consolidated $R/r^\prime$-band optical light curve (Figure~\ref{fig:model-FSRS}) from $3.0$ to $15.2$ days can be fit with a broken power law, breaking from $\alpha_r=-0.65\pm0.1$ to $-2.5\pm1.0$ at $10.8\pm1.4$ days. The shallowest possible decay of $\alpha=(3-3p)/4$, afforded by a uniform-density\footnote{For arbitrary $k$,  $\alpha = -[3(p-1)/4+k/(8-2k)]$ for $\nu_{\rm m}<\nu_{\rm obs} < \nu_{\rm c}$ and $\alpha=(2-3p)/4$ for $\nu_{\rm obs}<\nu_c$. In the latter case ($\nu_c<\nu_{\rm opt}$), the observed optical spectral index of $\beta\approx-1$ implies $p\approx2$ and predicts $\alpha\approx-1$, inconsistent with the observed shallow optical decline. In the former case ($\nu_X<\nu_c$), the spectral index implies $p\approx3$ and requires $k\approx10$, where the underlying \cite{bm76} solution breaks down.} environment ($k=0$) in the regime $\nu_m<\nu_{\rm opt}<\nu_c$, nevertheless still implies an unusually low value of $p=1.8\pm0.1$ and predicts an optical spectral index of $\beta = (1-p)/2\sim-0.4$, which is ruled out by steeper $\beta_{\rm opt}=-0.9\pm0.1$ inferred from a power-law fit to the $B$-band to $H$-band photometry at $\approx3.9$~days. 

The X-ray light curve evolves from $\alpha_{\rm X,1}\approx-0.3$ to $\alpha_{\rm X,2}\approx-1.6$ at $(8\pm2)$~days \citep{gcn44095}. The optical and X-ray decay rate from $3$--8 days appear consistent within the uncertainties. Extrapolating the $B$- to $J$-band optical spectral energy distribution (SED) at $\approx3.6$~days (derived from \citealt{gcn43993,gcn44000,gcn44002,gcn44003,gcn44006}) to the X-rays gives a flux consistent with the X-ray light curve at this time.  Furthermore, the X-ray spectral index\footnote{We note that EP/FXT finds a harder spectral index at $\approx2.5$~days, indicating either an evolution in $\beta_{\rm X}$ with time or a systematic effect. Since the NuSTAR observation spans a wider frequency range (3--79\,keV, we proceed with the latter in our analysis.}, $\beta_{\rm X}=-0.76\pm0.22$ \citep{gcn44063} is consistent with $\beta_{\rm opt}$. Together, these three observations indicate that the optical and X-rays are on the same synchrotron segment.

The quasi-monochromatic optical/X-ray break at $\approx10$~days and steep post-break decay is consistent with a jet break ($\alpha\approx-p$), suggesting $p\approx2.6$. This value of $p$ yields a spectral slope, $\beta\approx-0.8$ in the regime $\nu_{\rm X}\lesssim\nu_c$, which is roughly consistent with both the optical and X-ray spectral indices. On the other hand, the shallow pre-break optical and X-ray decline points to either an unusual hydrodynamic evolution (e.g., energy injection) or viewing angle effects (e.g., a structured jet viewed off-axis) rather than different origins\footnote{The X-ray light curve exhibits a late-time re-brightening (beginning within 10--27~days; \citealt{gcn44234,gcn44278}), which may be evidence for shock interaction or a separate emission component.} for the optical and X-ray emission. 

We find that an FS model with $p_{\rm FS}\approx2.6$, $\epsilon_{\rm e}\approx0.5$, $\epsilon_{\rm B}\approx1.5\times10^{-3}$, $A_*\approx0.1$, $E_{\rm K,iso}\approx7.2\times10^{52}$~erg, and $\theta_{\rm jet}\approx8^\circ$ with energy injection periods from $\approx0.1$--1~day ($E\propto t$) and 1.4--9~days ($E\propto t$) matches the X-ray and optical light curves well (Figure~\ref{fig:model-FSRS}). This model has $F_{\nu,\rm m, FS}\approx15.5$\,mJy, $\nu_{\rm c,FS}\approx2.7\times10^{17}\,{\rm Hz}$ (between the EP/FXT and NuSTAR bands), $\nu_{\rm m,FS}\approx1.2\times10^{12}<\nu_{\rm opt}$, and $\nu_{\rm sa,FS}\approx2.8\times10^{7}\,{\rm Hz}<\nu_{\rm radio}$ at 9~days, consistent with the soft X-ray and optical being on the same synchrotron segment. 

In this model, the prompt efficiency relative to the energy of the entire outflow is tiny, $\eta_{\gamma}\approx0.05\%$. Events exhibiting energy injection signatures are known to yield lower prompt efficiencies \citep[e.g.,][]{lbm+15}. If the energy injection is due to shells of low-Lorentz factor ejecta catching up and depositing energy into the FS \citep{rm98,sm00}, then this low prompt efficiency can be understood in a framework where the prompt emission (and hence, $E_{\gamma,{\rm iso}}$) is dominated by emission from the fast-moving ejecta while the value of $E_{\rm K,iso}$ inferred from the afterglow modeling is sensitive to the total kinetic energy, including that residing in slower moving ejecta. For this FS model, the energy prior to the start of injection is $E_{\rm K,iso}\approx\times10^{51}$~erg, which corresponds to a very reasonable value of $\eta_{\gamma}\approx26\%$, further supporting the energy injection scenario. 

\subsection{Afterglow Evolution -- Radio}
This FS model underpredicts the majority of the radio observations at $\gtrsim16$~days (Figure~\ref{fig:model-FSRS}). Similar excess radio emission has frequently been ascribed to the RS  \citep{kfs+99,sr03,kz03,bsfk03, lbz+13,pcc+14,vdhpdb+14,lab16,lag+19,lves+19,brf+23,gg23}. The observed evolution of the peak flux ($F_{\rm peak}\sim t^{-1.5}$) and peak frequency ($\nu_{\rm peak}\sim t^{-0.6}$) of the radio component is slower than expected from a standard RS model, an effect plaguing several previous RS studies \citep{lbz+13,pcc+14}. Following the spectacular failure of standard RS models for GRB\,221009A \citep{lam+23,brf+23,rvdhb+24}, an RS in a jet with angular structure was invoked to explain the radio emission in that event \citep{gg23,zwz24}. We implement the analytical model of \cite{zwz24} (Rohde et al., in prep\footnote{\url{https://github.com/rohdog2003/rsjetstruct}}) and present a structured-jet RS model with a deceleration time of $t_{\rm dec}\approx4.3$~days, which roughly matches the radio SEDs, in Figure~\ref{fig:model-FSRS}. 

This model has $F_{\nu\rm m,RS}\approx133\,{\rm mJy}$, $\nu_{\rm sa,RS}\approx30\,{\rm GHz}$, $\nu_{\rm m,RS}\approx8\,{\rm GHz}$, $\nu_{\rm c,RS}\approx1.4\times10^{14}\,{\rm Hz}$, $p_{\rm RS}\approx2.5$, $t_{\rm jet}\approx19$~days (consistent with the FS), $g\approx1.2$, and $k_\epsilon\approx2$ where $\Gamma_{{\rm ej}}\propto R^{-g}$ parametrizes the post-deceleration evolution of the shocked ejecta and $E_{{\rm k,iso}}(\theta)\propto\theta^{-k_\epsilon}$ parametrizes the angular distribution of the ejecta isotropic equivalent kinetic energy.

For completeness, we consider an alternate scenario where the radio flux is dominated by FS-like emission in Appendix~\ref{appendix:radioFS}. We find that any viable FS model that matches the radio emission will underpredict the X-ray and optical light curves, requiring a separate emission component and equal or greater complexity than the FS+RS model discussed above. These models may be distinguishable via future polarimetry observations that measure the time dependence of $\Pi_L$.

\newcommand{\fw}{0.32}

\section{Discussion}\label{sec:disc}
Our polarization observations offer two independent novel probes of the magnetic structure of the jet and its environment: $\Pi_0$ and RM. We now discuss possible interpretations of these measurements.

\subsection{Polarimetric Constraints on Magnetic Field Coherence Scale}
\label{sec:Pi0}

Our afterglow model indicates that the radio emission is entirely dominated by the RS component at the time of our radio polarimetric measurement (19.2~days). In the presence of global ordered fields, the maximum fractional polarization fraction of synchrotron radiation in the optically thin regime ($\nu_{\rm obs}>\nu_{\rm sa}$),$\Pi_{\rm thin}=(p+1)/(p+7/3)\approx0.72$ for $p\sim2.5$ (\citealt{pac70,rl86}; see also Appendix~\ref{appendix:PiSSA}). As we discussed above, the decrease in $\Pi_L$ close to and below the peak is consistent with SSA depolarization (Section~\ref{sec:vla} and Appendix~\ref{appendix:PiSSA}) and an intrinsic polarization $\Pi_0=(3.2\pm0.2)\%$, implying that the net polarization we observe on the optically thin segment is much less than the theoretical maximum of $\Pi_{\rm thin}$. 

Following \cite{lag+19}, we consider a model in which the observed polarization arises from the combined emission of intrinsically polarized, mutually incoherent regions, each characterized by a magnetic field that is ordered over a typical angular scale, $\theta_{\rm B}$ \citep{gk03,no04,gt05}. The number of patches contributing to the observed emission at time $t$ is $N\sim(\Gamma_{\rm ej}(t)\theta_{\rm B})^{-2}$, where $\Gamma_{\rm ej}(t)\approx\Gamma_{\rm sh}(t)(t/t_{\rm dec})^{-[g-(3-k)/2]/[(4-k)(2g+1)]}\propto t^{-0.1}$ (for $g=1.2$ and $k=2$) is the ejecta Lorentz factor \citep{kob00,gt05}. In our FS+RS model, $\Gamma_{\rm sh}\approx5.4$ at $19.2$~days, implying $\Gamma_{\rm ej}\approx4.7$ at that time. 

The observed polarization from $N$ patches is a random walk of $N$ steps in the $QU$ plane, such that $\Pi_{\rm L,obs}/\Pi_{\rm thin}\sim 1/\sqrt{N}$. Assuming a maximum intrinsic polarization of 72\% for $p\sim2.5$, this corresponds to $N\approx500$ and implies $\theta_{\rm B}\sim10^{-2}$. This measurement has a statistical uncertainty of $\approx10\%$, but larger systematic uncertainties from the RS dynamics and the stochasticity of the 2D random walk. This value of $\theta_{\rm B}$ is an order of magnitude larger than $\theta_{\rm B}\approx10^{-3}$ inferred from the ALMA polarimetry measurement of GRB\,190114C \citep{lag+19}; however, in that case the observation took place at $\approx7\times10^{-2}$ days in the rest frame and at $t\ll t_{\rm jet}$, whereas our VLA polarimetric observations of GRB~260310A are much later at $16.6$~days in the rest frame and at a time comparable with the jet-break, when considerable sideways spreading may take place \citep{rho99}. 

In Section~\ref{sec:vla}, we interpreted the decrease of $\Pi_L(\nu)$ with decreasing frequency as arising from SSA (Figure~\ref{fig:RMQU}). Combined with the above patchy shell model, this predicts that the polarization should rise again below $\nu_{\rm sa}$ to a fraction $(p+7)(p+1)^{-1}(6p+13)^{-1}\approx0.13$ (for $p=3.1$; see Appendix \ref{appendix:PiSSA}) of its value ($\Pi_0$) on the optically thin segment. For $\Pi_0=3.2\%$, this corresponds to $\approx0.4\%$, just beyond the reach of our existing observations. Furthermore, the polarization position angle is expected to rotate by $90^{\circ}$ across $\nu_{\rm sa}$. Lower frequency polarimetric observations may test this framework further. 

We note that optical polarimetric observations of GRB 260310A at 2.89 days post-trigger yielded a non-detection with an upper limit of $\Pi_L < 1.5\%$ \citep{gcn43990}. While dust along the LOS can depolarize optical emission, the Galactic extinction toward GRB 260310A is very low and the transient is significantly offset from the center of its host galaxy (Figure \ref{fig:host}), suggesting minimal dust column from either the Milky Way or the host interstellar medium (ISM). We therefore do not expect dust-induced depolarization to significantly affect the optical polarization measurement for this event.
Since SSA and Faraday depolarization are not relevant at optical wavelengths, we expect $\Pi_{\rm L,opt}\approx\Pi_{\rm L,radio}$ if the radio and optical emission are spatially co-located. 
In our FS+RS model, the optical emission is dominated by the FS at $\sim3$ days; since the FS is not expected to be strongly polarized \citep{mkca+13,gkg21}, this provides a naturally explanation for the non-detection of optical polarization in this event.
On the other hand, in any plausible models where the optical and radio emission arise from the same population of non-thermal electrons, the optical limit of $\Pi_{\rm L,opt}<1.5\%$ at 2.89 days relative to $\Pi_{\rm L,radio} \approx 3\%$ at 19.2 days would imply genuine temporal evolution of the polarization and require the magnetic field structure along the LOS to change over this period.

\subsection{Constraints on Faraday Depolarization and Origin of the high RM}
The high rotation measure, ${\rm RM_{int}}\approx-8.3\times10^3~{\rm rad\,m^{-2}}$, corresponds to a large rotation of $\chi$. %, $\Delta\chi\approx1$\,rad even at 25\,GHz. 
If the Faraday screen were co-located with the emitting region, we would expect strong Faraday depolarization across the source due to the differing paths of photons through the jet; however, the expected resulting deviations from $\chi\propto\lambda^2$ are not observed (Figure~\ref{fig:RMQU}). Instead, the observed frequency dependence of $\Pi(\nu)$ is well-explained by SSA depolarization, itself expected given clear SSA signatures in the radio SEDs (Figure~\ref{fig:model-FSRS}). This suggests a Faraday screen external to the emitting region, and allows us to quantify the coherence of such a screen. In the presence of stochastic RM variations of strength $\sigma_{\rm RM}$, we expect $\Pi\sim\Pi_0e^{-2\sigma_{\rm RM}^2\lambda^4}$ \citep{b66}. The absence of strong Faraday depolarization at $11$\,GHz implies $2\sigma_{\rm RM}^2\lambda^4\ll1$ at that frequency, requiring $\sigma_{\rm RM}\lesssim10^3\,{\rm rad\,m^{-2}}$. Relative to the observed RM, this implies a Faraday screen that is coherent at $\approx15\%$. We caution that this estimate is weakened if some of the observed drop in $\Pi$ at $\lesssim11$\,GHz is instead due to Faraday depolarization (we discuss this scenario further in Appendix~\ref{appendix:radioFS}). Further observations with a broader wavelength coverage may help disentangle these possibilities. 

In the following, we proceed under the assumption that the Faraday screen is external to the emitting region, and explore the physical origin of the large observed RM by assessing several plausible scenarios for the magnetized medium responsible for the Faraday rotation. Following the parameterization described in \citet{gt05}, we can relate the RM to physical quantities through the expression:
\begin{equation}
    \text{RM} = \frac{812}{(1+z)^2} \int_{0}^{L} \left(\frac{n_e}{\rm{cm^{-3}}}\right) \left(\frac{B_\parallel}{\rm{mG}}\right) \left(\frac{dl}{\rm{pc}}\right)~\rm{rad/m^2} \label{eq:rm}
\end{equation}
where $L$ is the distance from the source to the end of the magnetized medium, $n_e$ is the free electron number density, $B_\parallel$ is the magnetic field strength along the LOS, and $dl$ is the incremental LOS path length from the source to the observer.

First, we consider the case where the radio emission is produced in the RS and the RM is produced by the shocked CSM behind the FS. At the time of our polarimetric measurements (19.2~days), our FS+RS model yields $\Gamma_{\rm FS}\approx5.4$, shock radius $R_{\rm sh}\approx10^{19}$\,cm, and post-shock magnetic field, $B\approx1.4$\,mG. The density of the CSM at $R_{\rm sh}$ is $n_{\rm ext}\approx3\times10^{-4}\,{\rm cm}^{-3}$, implying a post-shock density, $n^\prime\approx4\Gamma_{\rm FS}n_{\rm ext}\approx5.8\times10^{-3}\,{\rm cm}^{-3}$. Approximating the shocked interstellar medium (ISM) as a shell of proper thickness, $R\sim R_{\rm sh}/\Gamma_{\rm sh}\sim2\times10^{18}$\,cm, we estimate ${\rm RM}_{\rm FS}\approx3\,{\rm rad\,m^{-2}}$. This is much smaller than the observed value of $|\rm{RM}|\approx6.3\times10^3\,{\rm rad\,m^{-2}}$, implying that the shocked CSM behind the FS cannot be responsible for the observed Faraday rotation. 

The ISM of the host galaxy could, in principle, contribute significantly to the observed RM. In the Milky Way, sight lines at low Galactic latitudes through the dense, magnetized ISM of the disk can produce RMs in excess of $\sim100-1000~\rm{rad\,m^{-2}}$ \citep{ojr12}. Pulsar studies have shown Galactic LOS magnetic field strengths to be of order a few $\mu\text{G}$ \citep[e.g.,][]{rk89,os93,hfm04,hmg11}. From Figure \ref{fig:host}, we can see that GRB 260310A is well offset from the center of its host galaxy, placing it in a region where the ISM column density is expected to be substantially lower than along sight lines through the inner disk. We therefore suspect that the host galaxy ISM is an unlikely origin for the bulk of the observed RM, assuming the host galaxy shares a similar Faraday rotation map to that of the Milky Way.

An alternative explanation for the large rotation measure is the presence of a dense, magnetized region local to the GRB progenitor. One such environment is an HII region, where the massive stellar progenitors to SN Ic-BL associated with LGRBs are thought to originate \citep{zy07}. The intense UV radiation and stellar winds from the massive progenitor can ionize and dynamically shape the surrounding medium, driving turbulent interactions within the ambient gas and dust \citep{zy07}. This process enhances the free electron density and magnetic field strength, naturally giving rise to elevated Faraday rotation measures \citep[exceeding several hundred $\rm{rad/m^2}$ in Galactic HII regions;][]{hmg11}. 

Modeling of the X-ray afterglows of long-duration GRBs has shown that their progenitors reside within dense regions of photoionized gas that span sizes of $5-100~\rm{pc}$ and free electron densities of $n_e \sim 10^2-10^4~\rm{cm^{-3}}$ \citep{tpl2026}. Independently, studies of magnetic field strengths in Galactic HII regions indicate that $B$ fields in these environments can range from $\sim 1-100~\mu\rm{G}$ (e.g., \citealt{hct81, hmg11, dcd21}). For these parameters, the expected RM spans from $\sim 4 \times 10^2$ to $\sim 8 \times 10^{7}~\rm{rad~m^{-2}}$, comfortably encompassing the observed value of $|\rm{RM_{int}}| \approx 8.3 \times 10^3~\rm{rad~m^{-2}}$. Our observed RM is therefore fully consistent with the population of HII regions previously associated with long GRBs and supports a massive star progenitor for GRB 260310A.
If the RM is predominately produced in an HII region surrounding the GRB, we would expect it to remain constant in time. Future multi-frequency polarimetry observations can thus test this scenario, and distinguish it from alternate GRB jet models that might allow some RM contribution from within the jet itself.

\section{Conclusions}\label{sec:conc}
We report the detection of linear polarization in GRB 260310A, marking the first centimeter polarization detection of a GRB and the first measurement of Faraday rotation in a GRB. We detect linearly polarized emission $11 - 25$ GHz, with the polarization fraction rising monotonically from $\Pi_L = (0.69 \pm 0.22)\%$ at 11 GHz to $\Pi_L = (3.18 \pm 0.18)\%$ at 25 GHz. The depolarization toward lower frequencies is likely dominated by SSA, which reduces the polarized intensity as the emitting region becomes optically thick.

Our observations reveal a frequency-dependent variation in the polarization angle consistent with Faraday rotation with $\text{RM} = -6250 \pm 70~\text{rad m}^{-2}$ ($\rm{RM_{int}}=-8300 \pm 90~\text{rad m}^{-2}$ at the GRB redshift). The magnitude of this RM is consistent with linearly polarized emission propagating through a dense HII region, consistent with a massive star progenitor for GRB 260310A and pointing to a highly magnetized intervening medium. 

These results demonstrate that GRB afterglows are linearly polarized across a broad range of radio frequencies, and that their polarization properties encode information about both the intrinsic jet magnetization and the intervening magnetized medium. Future multi-frequency polarimetric monitoring campaigns spanning days to weeks will trace the co-evolution of magnetic field geometry and afterglow emission, offering new constraints on the structure and dissipation of magnetic fields in GRB afterglows.

\section*{Acknowledgments}
The National Radio Astronomy Observatory (NRAO) is a facility of the National Science Foundation operated under cooperative agreement by Associated Universities, Inc. Some observations reported here were obtained at the MMT Observatory, a joint facility of the University of Arizona and the Smithsonian Institution. K.~D.~A. gratefully acknowledges support from the Alfred P. Sloan Foundation. N.~F. acknowledges support from the National Science Foundation Graduate Research Fellowship Program under Grant No. DGE-2137419.

This work made use of the following software packages: \texttt{astropy} \citep{astropy:2013,astropy:2018,astropy:2022}, \texttt{Jupyter} \citep{2007CSE.....9c..21P}, \texttt{matplotlib} \citep{Hunter:2007}, \texttt{numpy} \citep{numpy}, \texttt{pandas} \citep{mckinney-proc-scipy-2010}, \texttt{python} \citep{python}, \texttt{scipy} \citep{2020SciPy-NMeth}, and \texttt{emcee} \citep{emcee-Foreman-Mackey-2013}.

Software citation information aggregated using \texttt{\href{https://www.tomwagg.com/software-citation-station/}{The Software Citation Station}} \citep{software-citation-station-paper,software-citation-station-zenodo}.

% \bibliography{grb_alpha,sample701}{}
\bibliography{merged}{}
\bibliographystyle{aasjournalv7}

\appendix
\section{Polarization Calibration Verification and Systematic Checks}\label{sec:polsys}
\begin{figure}
    \centering
    \includegraphics[width = 0.5\textwidth]{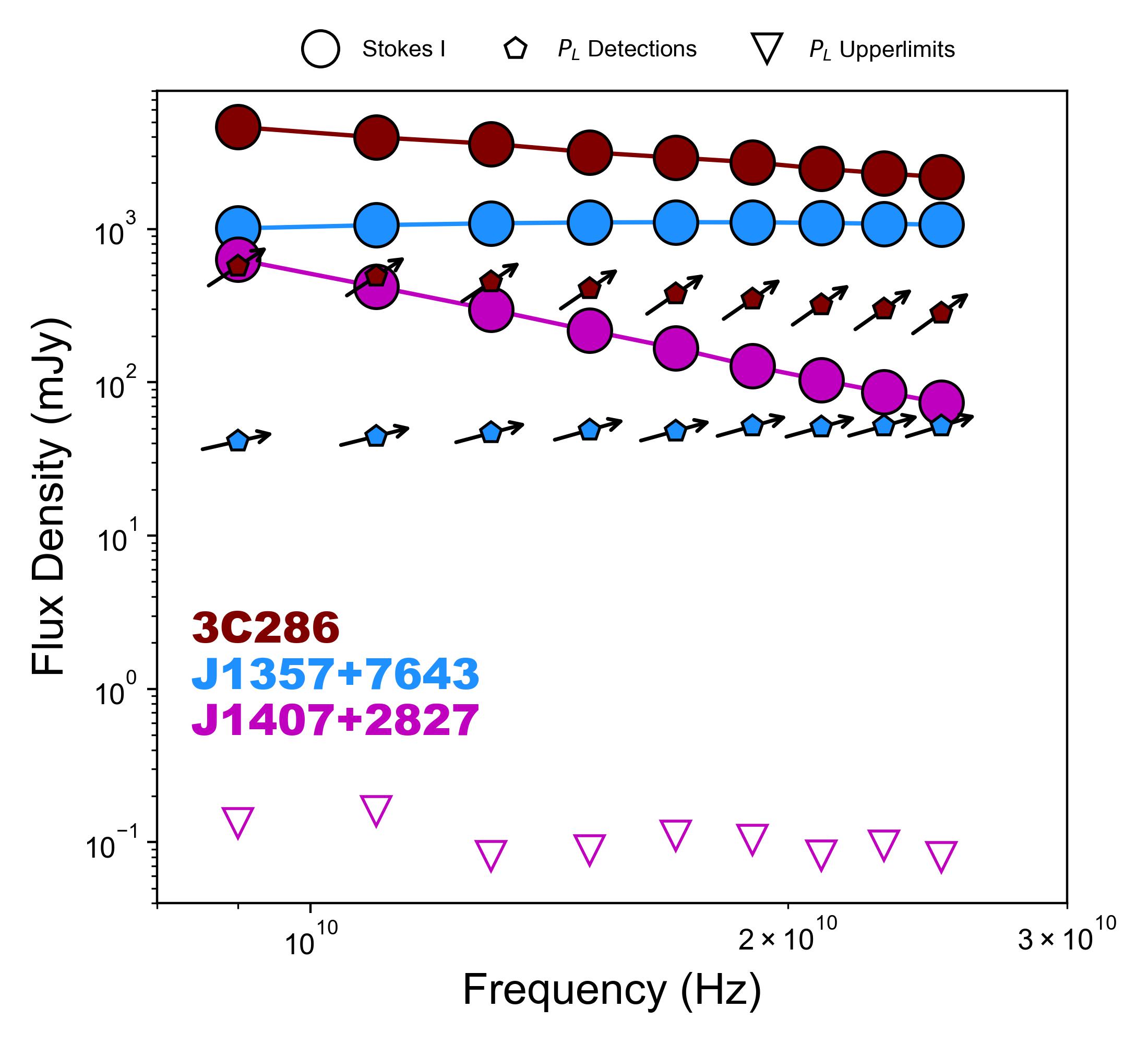}
    \caption{The spectral energy distribution for the bandpass, flux, and polarization angle calibrator 3C286, the gain calibrator J1357+7643, and the leakage calibrator J1407-2827 for the total intensity (Stokes $I$, circles) and linearly polarized intensity (Stokes $P_L$, pentagons). For the leakage calibrator, we show the 3$\sigma$ upper limits on Stokes $P_L$ (triangles). For the polarization flux densities we also overlay the direction of the polarization angle (with the angle measured from the horizontal axis).}
    \label{fig:cal_seds}
\end{figure}

\begin{figure*}
    \centering
    \includegraphics[width = \textwidth]{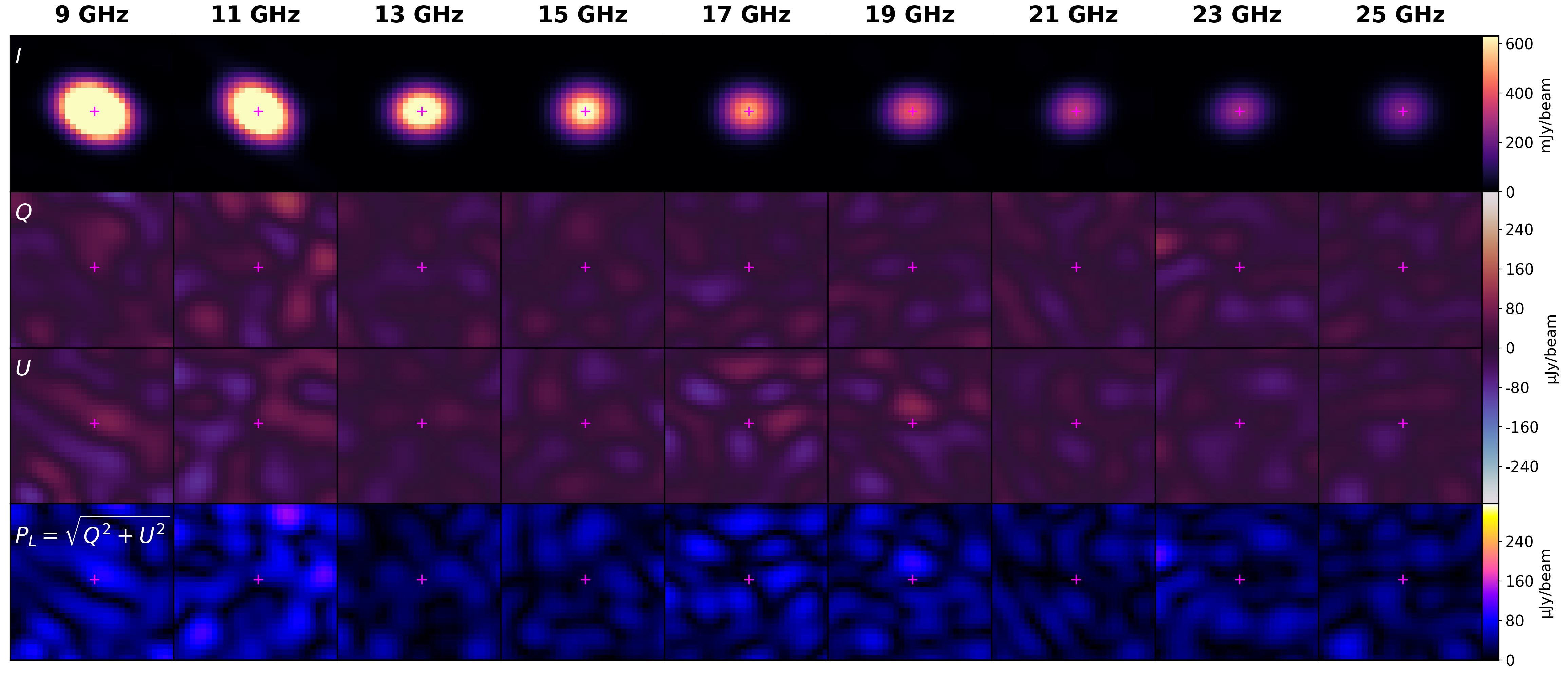}
    \caption{Same as Figure \ref{fig:pol_imgs} but for the leakage calibrator J1407+2827.}
    \label{fig:Lcal_imgs}
\end{figure*}

\begin{figure*}
    \centering
    \includegraphics[width = \textwidth]{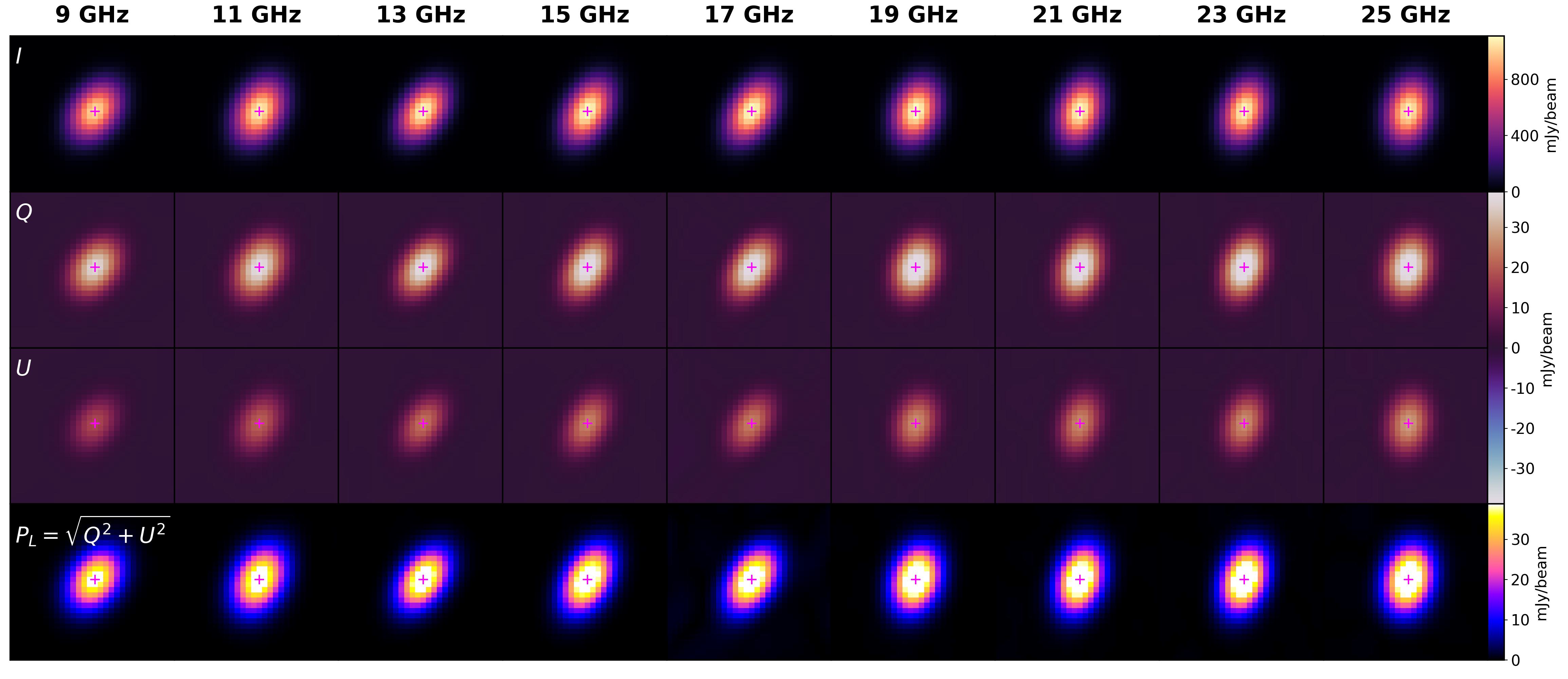}
    \caption{Same as Figure \ref{fig:pol_imgs} but for the phase calibrator J1357+7643.}
    \label{fig:Gcal_imgs}
\end{figure*}

To assess the robustness of our polarization calibration, we performed a series of diagnostic checks on each calibrator. We describe these checks below and present the resulting calibrator SEDs and images in Figures \ref{fig:cal_seds}, \ref{fig:Lcal_imgs}, and \ref{fig:Gcal_imgs}. 

During calibration, the absolute polarization angle and fraction were set using 3C286, with Stokes $I$ flux densities, polarization fractions, and polarization angles fixed to the frequency-dependent values tabulated by NRAO. To verify that this standard was faithfully transferred to the data, we imaged 3C286 across all observed frequencies and confirmed that the recovered polarization fractions and angles are consistent with the tabulated values to within the measurement uncertainties (Figure \ref{fig:cal_seds}).

We next imaged the leakage calibrator J1407+2827, a source selected for its negligibly low intrinsic polarization, to verify the quality of the instrumental polarization (D-term) calibration. As expected, we detect a strong point source in Stokes $I$ with no discernible emission in the Stokes $Q$ or $U$ images, which are instead consistent with noise (Figure \ref{fig:Lcal_imgs}). The $3\sigma$ upper limits on polarized emission from J1407+2827 correspond to $\lesssim 0.1\%$ of Stokes $I$, confirming that the instrumental leakage terms are properly calibrated to at least this level. For GRB 260310A, this leakage floor corresponds to $\sim$10 $\mu$Jy, which is comparable to the RMS noise in the Stokes $Q$ and $U$ target images and well below our detected polarization signal of $\sim60-270~\mu\rm{Jy}$ (Table \ref{tab:Stokes_pol}).

As an additional verification that the detected polarization is astrophysical and not an artifact of the calibration or imaging procedure, we confirmed that the linear polarization signal in the target field is present both before and after self-calibration. As described in Section \ref{sec:vla}, self-calibration was performed on Stokes $I$ (phase-only), with the derived phase corrections subsequently applied to all Stokes parameters. The Stokes $Q$ and $U$ flux densities measured in each band are consistent before and after the application of the self-calibration solutions, with changes well within the measurement uncertainties. 

We also imaged the phase calibrator J1357+7643, which was not used to calibrate the cross-hand (RL/LR) visibilities and therefore serves as an independent check on the polarization calibration. J1357+7643 exhibits weakly polarized emission at each observed frequency with a stable polarization angle across the band, consistent with a low rotation measure. The absence of significant Faraday rotation in this calibrator LOS is consistent with the modest Milky Way RM contribution expected at intermediate Galactic latitudes, further supporting our conclusion that the large RM observed toward GRB 260310A is intrinsic to the source or its local environment (Figure \ref{fig:Gcal_imgs}).

In Stokes $V$, we find that the images show low-level beam squint artifacts due to the offset in the RR and LL beam centers (common in off-axis feeds, which is the case for the VLA). This differential pointing introduces a systematic artifact in Stokes $V = (\text{RR} - \text{LL})/2$ that manifests as equal and opposite lobes at the position of the source in the Stokes $V$ image. In our data, this artifact appears at the level of $\sim$0.2\% of Stokes $I$ in only the bright calibrator sources. For the $\sim$10 mJy flux density observed in GRB 2610310A the same fractional level ($\sim$20 $\mu$Jy) falls below our linear polarization detections, and no beam squint artifact is detected in the target Stokes $V$ images. Beam squint introduces a differential gain error between the parallel-hand correlations (RR and LL) and therefore contaminates only Stokes $V$, leaving the cross-hand products that form Stokes $Q$ and $U$ unaffected \citep{uc08}. The linear polarization detection and rotation measure reported in this work are therefore not compromised by this low level beam squint.

\section{Frequency-Dependent Synchrotron Polarization with Self-Absorption}
\label{appendix:PiSSA}
Following \cite{pac70}, we derive a simple model for the frequency-dependent linear polarization fraction, $\Pi(\nu)$, of synchrotron emission in the presence synchrotron self-absorption. We start with the assumption of locally ordered magnetic fields and later relax this assumption to account for magnetic field anisotropies in the visible region. The model captures the transition between the optically thin and optically thick regimes and provides a convenient framework for forward modeling of Stokes parameters. 

For a power-law electron distribution $N(\gamma) \propto \gamma^{-p}$, the emission ($j_\nu$) and absorption ($\alpha_\nu$) coefficients of the two orthogonally polarized modes perpendicular ($\perp$) and parallel ($\parallel$) to the magnetic field projected on the sky are proportional to
\begin{align}
j_{\perp,\parallel}\propto 1\pm \frac{p+1}{p+7/3}\\
\alpha_{\perp,\parallel}\propto 1\pm \frac{p+2}{p+10/3}, 
\end{align}
where the $+$ sign corresponds to the $\perp$ mode and the $-$ sign to the $\parallel$ mode. We assume that the synchrotron self-absorption is dominated by the same non-thermal electrons responsible for the synchrotron emission. In the basis aligned with the polarization eigenmodes, the fractional linear polarization is 
\begin{align}
\Pi=\frac{Q}{I} = \frac{I_\perp-I_\parallel}{I_\perp+I_\parallel} 
= \frac{S_\perp(1-e^{-\tau_\perp}) - S_\parallel(1-e^{-\tau_\parallel})}
{S_\perp(1-e^{-\tau_\perp}) + S_\parallel(1-e^{-\tau_\parallel})},
\label{eq:fracpolofS}
\end{align}
where $S_{\perp,\parallel} \equiv j_{\perp,\parallel}/\alpha_{\perp,\parallel}$ is the source function and $\tau_{\perp,\parallel}\equiv\alpha_{\perp,\parallel}L$ is the optical depth (for a common path-length $L$) for each mode, respectively\footnote{This derivation assumes that the intrinsic Faraday rotation is small over a distance of one photon mean free path, i.e., $d\chi_F/ds\ll\alpha$, where $d\chi_F/ds$ is the Faraday rotation per unit path length and $\alpha$ is the mean absorption coefficient. See \cite{pac70} for a broader treatment including Faraday depolarization.}. 

In the optically thin limit, $I_\nu\rightarrow S_\nu\tau_\nu = j_\nu L$ and the fractional polarization, 
\begin{align}
\Pi_{\rm thin} = \frac{j_\perp-j_\parallel}{j_\perp+j_\parallel} = \frac{j_R-1}{j_R+1} = \frac{p+1}{p+\frac{7}{3}},
\label{eq:PiThin}
\end{align}
where
\begin{align}
j_R \equiv \frac{j_\perp}{j_\parallel} = \frac{1+\Pi_{\rm thin}}{1-\Pi_{\rm thin}} 
= \frac{3p+5}{2}.
\label{eq:jR}
\end{align}
In the optically thick limit, $I_\nu\rightarrow S_\nu$ and the fractional polarization, 
\begin{align}
\Pi_{\rm thick} = \frac{S_\perp-S_\parallel}{S_\perp+S_\parallel} 
= \frac{j_\perp/\alpha_\perp - j_\parallel/\alpha_\parallel}
       {j_\perp/\alpha_\perp + j_\parallel/\alpha_\parallel}
= \frac{j_\perp\alpha_\parallel - j_\parallel\alpha_\perp}{j_\perp\alpha_\parallel + j_\parallel\alpha_\perp} = \frac{j_R-\alpha_R}{j_R+\alpha_R}
= -\frac{3}{6p + 13},
\label{eq:PiThick}
\end{align}
where the negative sign indicates a $90^\circ$ rotation of the polarization angle relative to the optically thin case, and
\begin{align}
\alpha_R \equiv \frac{\alpha_\perp}{\alpha_\parallel} = \frac{1-\Pi_{\rm thick}}{1+\Pi_{\rm thick}}j_R = \frac{3p+8}{2}.
\label{eq:alphaR}
\end{align}

The total optical depth,
\begin{align}
\tau_\nu = \left(\frac{\nu}{\nu_{\rm sa}}\right)^{-(p+4)/2} = \frac{\tau_\perp + \tau_\parallel}{2},
\label{eq:tautotal}
\end{align}
where 
\begin{align}
\tau_\perp &=\frac{2\tau_\nu\alpha_\perp}{\alpha_\perp + \alpha_\parallel} 
= \frac{2\tau_\nu\alpha_R}{1+ \alpha_R} \label{eq:tauperp},\\ 
\tau_\parallel &=\frac{2\tau_\nu\alpha_\parallel}{\alpha_\perp + \alpha_\parallel} = \frac{2\tau_\nu}{1 + \alpha_R} \label{eq:tauparallel}, 
\end{align}
and $\nu_{\rm sa}$ is the synchrotron self-absorption frequency, which we define here by $\tau_\nu(\nu_{\rm sa}) = 1$. Equation~\ref{eq:fracpolofS} for the fractional polarization as a function of frequency in the presence of SSA can then be written as 
\begin{align}
\Pi(\nu) = \frac{\Pi_0}{\Pi_{\rm thin}}\frac{S_R(1-e^{-\tau_\perp}) - (1-e^{-\tau_\parallel})}
{S_R(1-e^{-\tau_\perp}) + (1-e^{-\tau_\parallel})},
\label{eq:fracpolofSR}
\end{align}
where $S_R\equiv j_R/\alpha_R$ and $\Pi_0<1$ is a free parameter that accounts for the fact that the observed polarization may be lower than the theoretically predicted maximum value due to, e.g., stochastic variations in $B$ in the source region contributing to observable emission (see Section~\ref{sec:Pi0}). Since $\tau_\perp/\tau_\parallel=\alpha_R=(3p+8)/2>1$, the perpendicular mode becomes optically thick first (at a higher frequency), where the parallel mode is still optically thin.  This differential optical depth effect leads to a null in $\Pi(\nu)$ where $I_\perp$ and $I_\parallel$ cancel at a frequency given by the solution of the expression
\begin{align}
    S_R(1-e^{-\frac{2\alpha_R\tau_\nu}{1+\alpha_R}})=1-e^{-\frac{2\tau_\nu}{1+\alpha_R}}. 
\end{align}

For a given value of $p$, we can calculate $\Pi_{\rm thick}$, $\Pi_{\rm thin}$, $j_R$ and $\alpha_R$ from equations~\ref{eq:PiThick}, \ref{eq:PiThin}, \ref{eq:jR}, and \ref{eq:alphaR}, respectively. Equations~\ref{eq:tautotal}, \ref{eq:tauperp}, and \ref{eq:tauparallel} together can be used to compute $\tau_\perp(\nu)$ and $\tau_\parallel(\nu)$. 
The total intensity in this framework (assuming the spectral peak is at $\nu_{\rm sa}$) is 
\begin{align}
    I(\nu) = S_\nu (1-e^{-\tau_\nu}) = I_0 \left(\frac{\nu}{\nu_{\rm sa}}\right)^{5/2}(1-e^{-\tau_\nu}), 
\end{align}
with $\tau_\nu$ given by equation~\ref{eq:tautotal}. Given the total intensity $I(\nu)$ and polarization fraction $\Pi(\nu)$, the Stokes parameters are
\begin{equation}
Q(\nu) = I(\nu)\,\Pi(\nu)\cos\left[2\chi(\nu)\right],
\end{equation}
\begin{equation}
U(\nu) = I(\nu)\,\Pi(\nu)\sin\left[2\chi(\nu)\right], 
\end{equation}
where the polarization angle is given by 
\begin{equation}
\chi(\nu) = \chi_0 + {\rm RM}\,\lambda^2,
\end{equation}
where $\lambda = c/\nu$ and RM is the rotation measure. The $90^\circ$ rotation across $\nu_{\rm sa}$ is captured by the sign change in $\Pi(\nu)$ and therefore does not need to be explicitly included in $\chi(\nu)$.

\section{Faraday Rotation from an SSA-peaked Relativistic Shock}
\label{appendix:radioFS}
As an alternative to the FS+RS model, we consider a scenario where the radio observations are dominated by FS-like emission. We find that a FS in a wind-like environment with $p\approx2.4$, $\epsilon_{\rm e}\approx0.1$, $\epsilon_{\rm B}\approx0.1$, $A_*\approx0.3$, $E_{\rm K,iso}\approx1.1\times10^{51}$~erg and $\theta_{\rm jet}\gtrsim40^\circ$, roughly matches the radio SEDs at $\approx16$--25 days.
This FS-only model has $F_{\nu,\rm m, FS}\approx31$\,mJy and $\nu_{\rm sa,FS}\approx4.7\times10^{7}\,{\rm Hz}<\nu_{\rm radio}<\nu_{\rm m,FS}\approx9.4\times10^{12}<\nu_{\rm c,FS}\approx5.5\times10^{13}\,{\rm Hz}<\nu_{\rm opt}$ at the time of our polarimetric observations (19.2~days). This model under-predicts the high-frequency (230\,GHz) radio data as well as the optical and X-ray emission, requiring a separate component at those frequencies. 

If the observed Faraday rotation were produced within the source, we can obtain an independent estimate of the magnetic field coherence scale by modeling the FS as a spherical shock traveling at Lorentz factor $\Gamma$ into an ambient medium with density $\rho_{\rm ext}(R)$ as a function of the shock radius, $R$. The image of the source has an extent $R_\perp$ transverse to the line of sight, corresponding to an observed angular radius of
\begin{equation}
\theta = \frac{R_\perp}{d_A},
\end{equation}
where $d_A$ is the angular diameter distance. In the following, we associate the spectral flux density at the synchrotron self-absorption (SSA) frequency ($\nu_{\rm sa}$) with black body emission, subsuming factors of order unity into a correction term, $\xi(p)$, such that the brightness temperature at $\nu_{\rm sa}$ in the co-moving frame is given by
\begin{equation}
3k_BT'_b \simeq \xi m_e c^2 \gamma_a. 
\end{equation}
The Doppler-boosted and redshifted brightness temperature is 
\begin{equation}
T_{b,\mathrm{obs}} = \frac{\delta}{1+z} T'_b, 
\end{equation}
where $\delta\sim\Gamma(1+\beta)$ is the Doppler factor. Taking the image to be a uniformly illuminated disk, the observed flux density is the product of the angular area multiplied by the intensity, 
\begin{equation}
F_a = \pi \theta^2 \frac{2\nu_{\rm sa}^2}{c^2} k T_{b,\mathrm{obs}}.
\end{equation}
Solving for the electron Lorentz factor yields,
\begin{align}
\gamma_a &=
\frac{3F_a d_A^2 (1+z)}
{2\pi \xi m_e \nu_{\rm sa}^2 R_\perp^2 \delta} \\ 
&= 19\,
{\xi}^{-1}
\left(\frac{F_a}{10.3\ \mathrm{mJy}}\right)
\left(\frac{\nu_{\rm sa}}{13\ \mathrm{GHz}}\right)^{-2}
\left(\frac{R_\perp}{1.5\times10^{17}\ \mathrm{cm}}\right)^{-2}
\left(\frac{d_A}{1.7\times10^{27}\ \mathrm{cm}}\right)^2 \nonumber
\left(\frac{\delta}{2.5}\right)^{-1}
\left(\frac{1+z}{1.153}\right).
\end{align}
Inverting the synchrotron characteristic frequency (assuming a pitch angle of $90^\circ$),
\begin{equation}
\nu(\gamma) = \frac{\delta}{1+z} \frac{3eB'}{4\pi m_e} \gamma^2,
\end{equation}
for the magnetic field, 
\begin{align}
B' &= \frac{4\pi m_e}{3e} \frac{(1+z)\nu_{\rm sa}}{\delta \gamma_a^2}\\
&=
4.1 \mathrm{G}\,
\xi^2
\left(\frac{F_a}{10.3\ \mathrm{mJy}}\right)^{-2}
\left(\frac{R_\perp}{1.5\times10^{17}\ \mathrm{cm}}\right)^{4}
\left(\frac{d_A}{1.7\times10^{27}\ \mathrm{cm}}\right)^{-4} 
\left(\frac{\delta}{2.5}\right)
\left(\frac{\nu_{\rm sa}}{13\ \mathrm{GHz}}\right)^{5}
\left(\frac{1+z}{1.153}\right)^{-1}.
\end{align}

For a strong relativistic shock, the proper post-shock number density is
\begin{equation}
 n'_{\rm ps} \simeq 4\Gamma n_{\rm ext}.
\end{equation}

The Faraday rotation measure for a uniform shell of proper thickness $\Delta R$ is
\begin{equation}
{\rm RM} =
\frac{e^3}{8\pi^2 \epsilon_0 m_e^2 c^3}
\int \frac{n'_e B'_\parallel  dl'}{(1+z)^2}
\simeq
\frac{e^3}{8\pi^2 \epsilon_0 m_e^2 c^3}
\frac{\eta_B n'_{\rm ps} B' \Delta R}{(1+z)^2}
\simeq
\frac{2.63\times10^{3}\eta_B }{(1+z)^2}
\frac{n'_{\rm ps}}{{\rm cm}^{-3}} \frac{B'}{\rm G} \frac{\Delta R}{10^{16}\,\rm cm} \ \mathrm{rad\ m^{-2}},
\end{equation}
where we have written $B'_\parallel = \eta_B B'$. Substituting the shock-jump density and the SSA-derived magnetic field, we have 
\begin{align}
{\rm RM}_{\rm obs} \simeq\;&
7.6\times10^{4}\,
\left(\frac{F_a}{10.3\ \mathrm{mJy}}\right)^{-2}
\left(\frac{\nu_{\rm sa}}{13\ \mathrm{GHz}}\right)^{5}
\left(\frac{1+z}{1.153}\right)^{-3} \nonumber\\
&\times
\eta_B\xi^2
\left(\frac{R_\perp}{1.5\times10^{17}\ \mathrm{cm}}\right)^{4}
\left(\frac{d_A}{1.7\times10^{27}\ \mathrm{cm}}\right)^{-4} 
\left(\frac{\delta}{2.5}\right)^2
\left(\frac{n_{\rm ext}}{0.15\ \mathrm{cm^{-3}}}\right)
\left(\frac{\Delta R}{10^{16}\ \mathrm{cm}}\right) 
\ \mathrm{rad\ m^{-2}}, 
\end{align}
where we have scaled the values to the image size and external medium density predicted for the FS using the \cite{bm76} solution at 19.2~days following the formalism of \cite{gs02}. Relative to the measured ${\rm RM}\sim6.3\times10^3\,{\rm rad\,m^{-2}}$, this implies $\eta_B\sim0.1$. Finding $\eta_B<1$ implies that Faraday rotation in a FS-produced $B$-field may generate the observed RM in this model. However, we note that this conclusion is strongly sensitive to the image size ($R_\perp$) and to the unknown thickness of the region responsible for the Faraday rotation, $\Delta R$. 

\end{document}